\documentclass[11pt]{article}
\usepackage{moriond,epsfig}

\def\be{\begin{equation}}
\def\ee{\end{equation}}
\def\bea{\begin{eqnarray}}
\def\eea{\end{eqnarray}}

\begin{document}
\vspace*{4cm} \title{INTRODUCTION TO EXTRA DIMENSIONS~\footnote{Based
on a talk given at the 41st Rencontres de Moriond on Electroweak
Interactions and Unified Theories, La Thuile, Aosta Valley, Italy,
11-18 March 2006.}}

\author{M. QUIROS}

\address{Instituci\'o Catalana de Recerca i Estudis Avan\c{c}ats
(ICREA)\\ and\\ Theoretical Physics Group, IFAE/UAB\\ E-08193
Bellaterra (Barcelona) Spain}

\maketitle\abstracts{ The aim of this talk is to provide non-experts
with a brief and elementary introduction on the field of extra
dimensions. The main motivation for extra dimensions relies on the
more fundamental string theories that predict ten (or eleven)
space-time dimensions. Extra dimensions must be compactified and there
appear branes where gauge and/or gravity propagates. Compactification
relates string constants (string scale and string coupling) with
four-dimensional constants (Planck scale and gauge coupling). Only
gravity can propagate in dimensions transverse to the brane. They can
be detected either by gravitational (table-top) or by collider
experiments where Kaluza-Klein graviton production appears as missing
energy. Transverse dimensions can be as large as the
sub-millimeter. Ordinary matter can also propagate in dimensions
parallel to the brane. It can give rise to bumps in the dilepton
invariant mass in hadron colliders or contribute by indirect effects
to the electroweak observables. Longitudinal dimensions can be probed
at LHC up to a scale of 6.7 TeV (9 TeV) for one (two) extra
dimension(s). Extra dimensions also give rise to new theoretical ideas
related to supersymmetry and electroweak breaking. Some of these ideas
are reviewed.}

\section{Introduction}
These notes contain very elementary and introductory material on the
field of (large) extra dimensions mainly addressed to non-expert
theorists and experimentalists wishing to initiate themselves in the
subject. Although this author has contributed to some of the original
developments that will be reviewed, no effort will be made for
originality here.

Extra dimensions were introduced to solve classical problems of
Particle Physics. Mainly to solve the hierarchy problem. Here two main
avenues were open.

$\bullet$ If there is a ``warped'' extra dimension, large scales at
the Planck brane are redshifted at the TeV brane~\cite{Randall:1999ee}
as it is shown in the cartoon of Fig.~\ref{fig1} (left panel). The
(RS) relationship between the weak and Planck scales is given by

\begin{equation}
M_W=e^{-2 k \pi R}M_{P\ell}
\end{equation}

$\bullet$ If there are $n$ very large extra dimensions where {\it only gravity
propagates} the Planck scale is ``reduced'' by the large
compactification volume $V\sim R^n$. In this case the (ADD) relation
between weak and Planck scales is~\cite{Arkani-Hamed:1998rs}

\begin{equation}
M_{W}\simeq\left[M_{Pl}\,R\right]^{-\frac{n}{n+2}} M_{Pl}
\end{equation}

\noindent This solution is illustrated in the right panel of
Fig.~\ref{fig1}.

\begin{center}
\hskip 2cm
\begin{figure}[htb]\hspace{1.2cm}
\psfig{figure=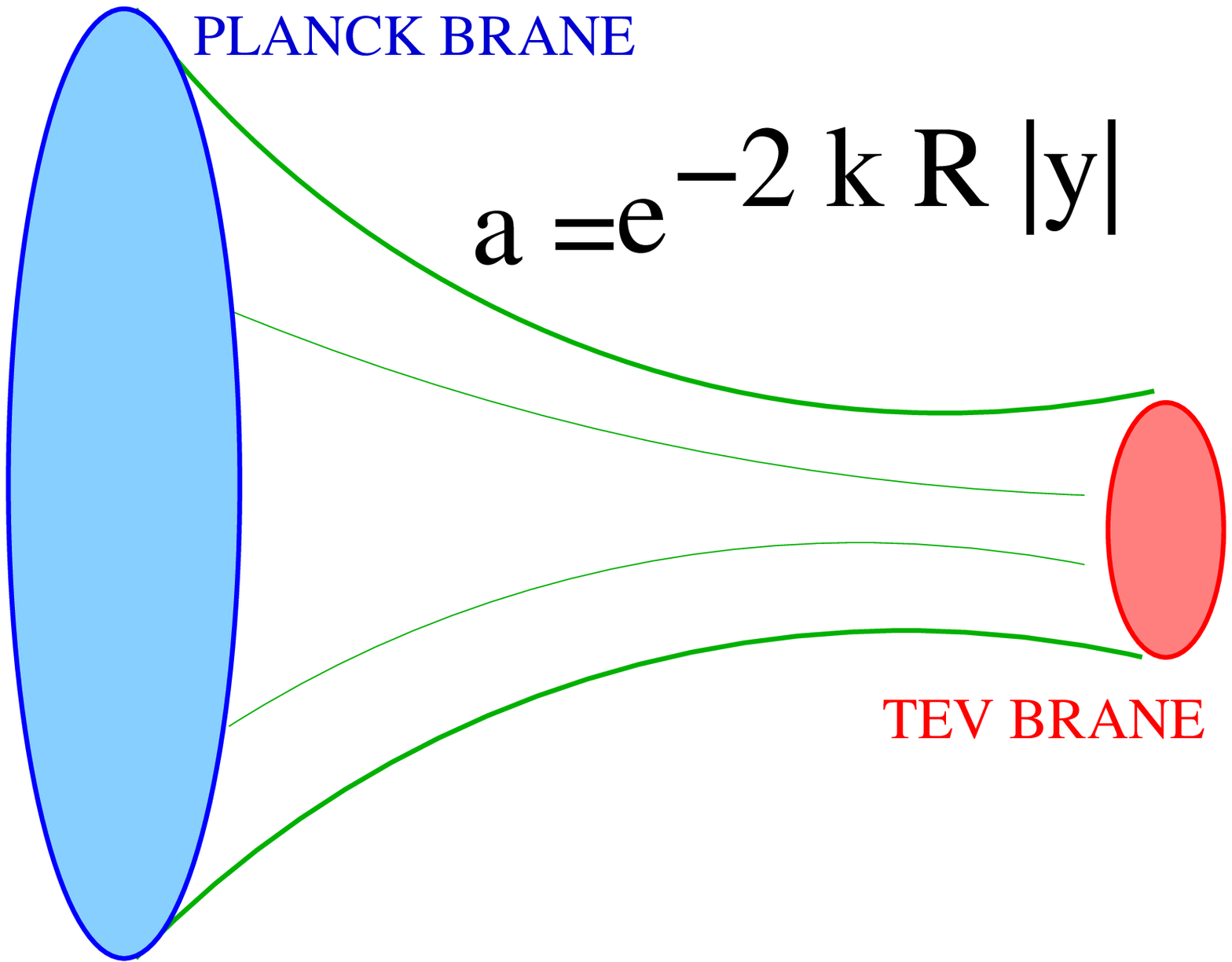,height=1.5in}
\hspace{1cm}\psfig{figure=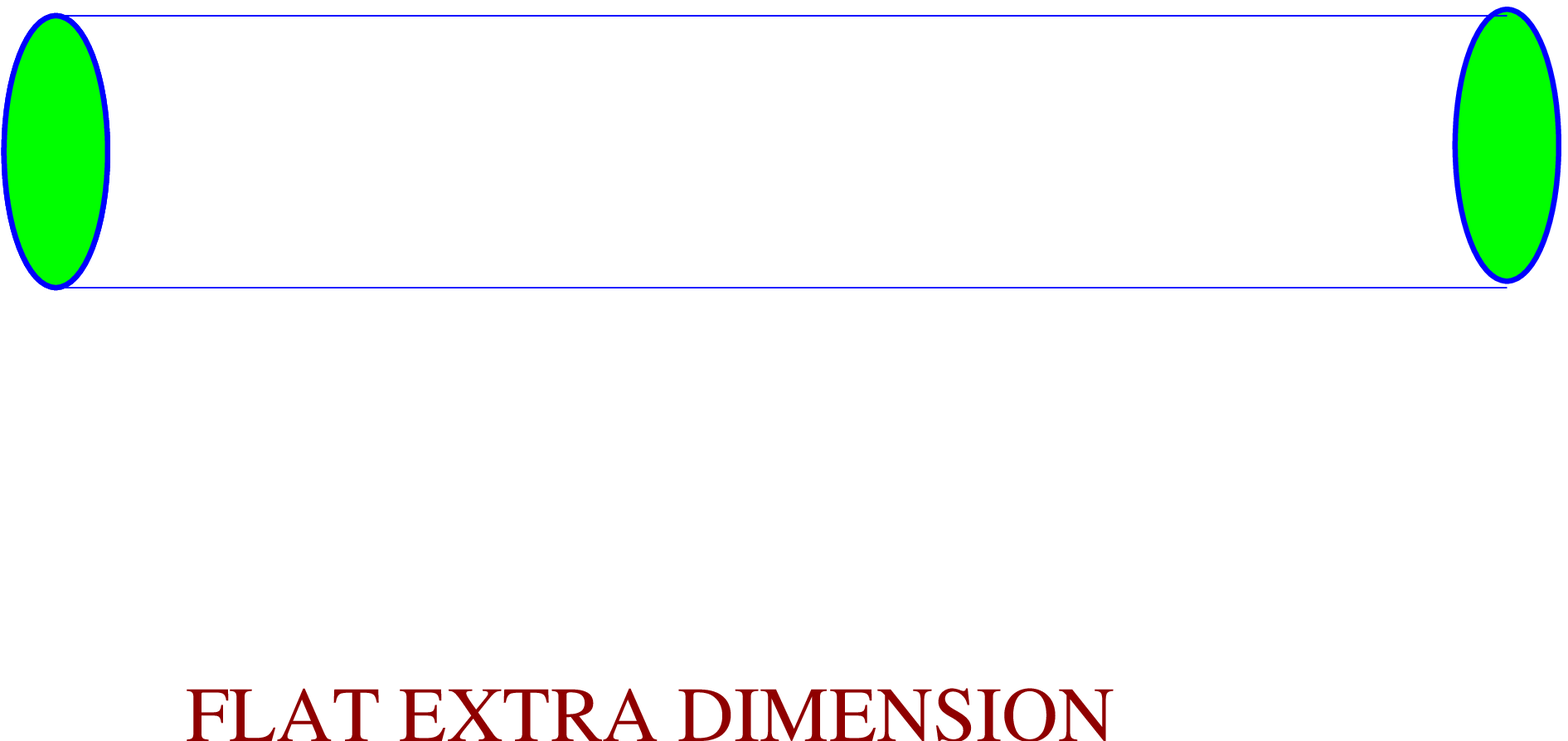,height=1.5in}
\caption{Left panel: The RS solution to the hierarchy; $a$ is the warp
factor, $k$ is the AdS curvature and $R$ the compactified radius.
Right panel: The ADD solution to the hierarchy.
\label{fig1}}
\end{figure}
\end{center}

\section{Where do extra dimensions come from?}

\begin{center}
\begin{figure}[htb]
\hspace{2cm}\psfig{figure=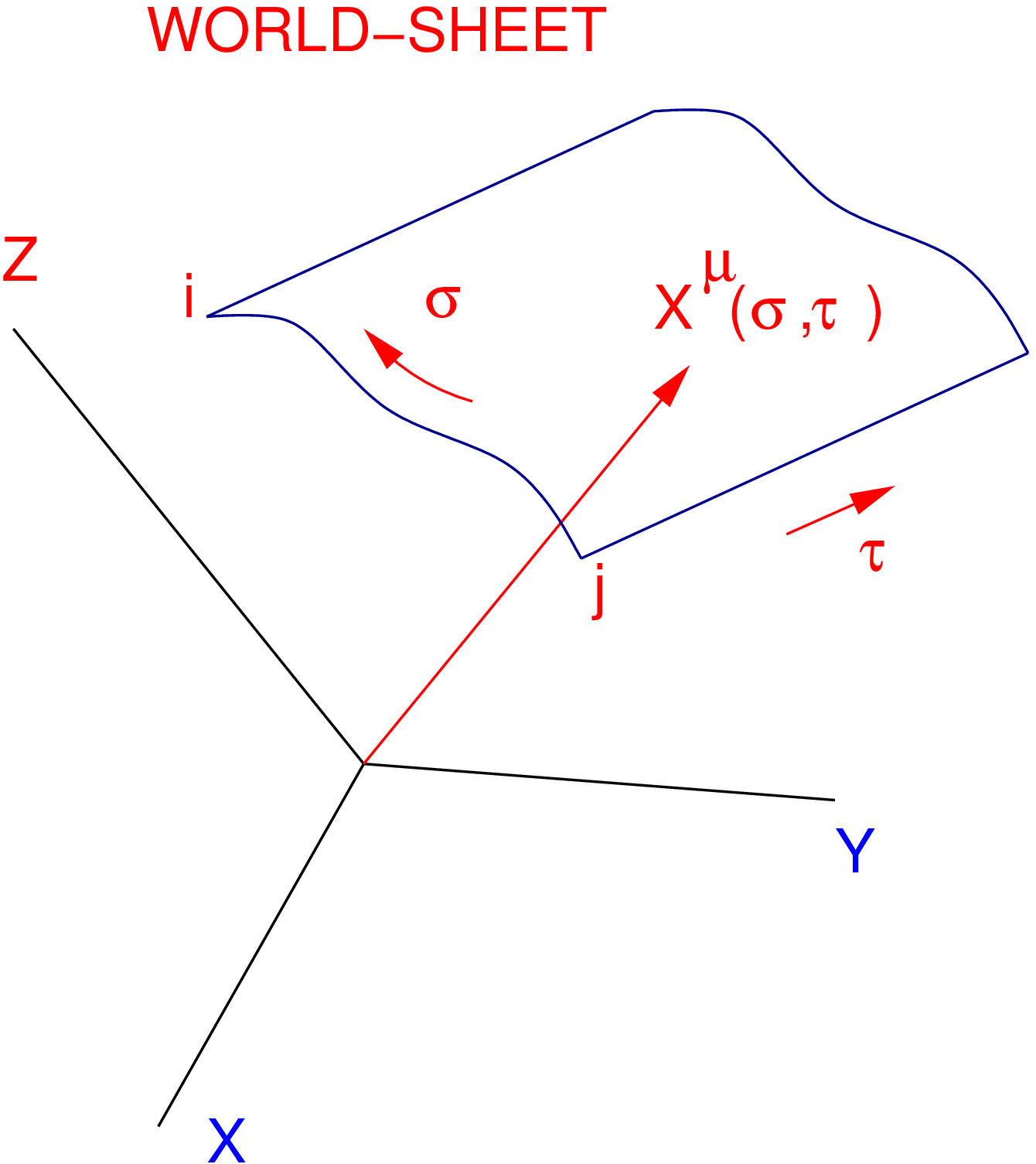,height=2.5in}
\hspace{1cm}\psfig{figure=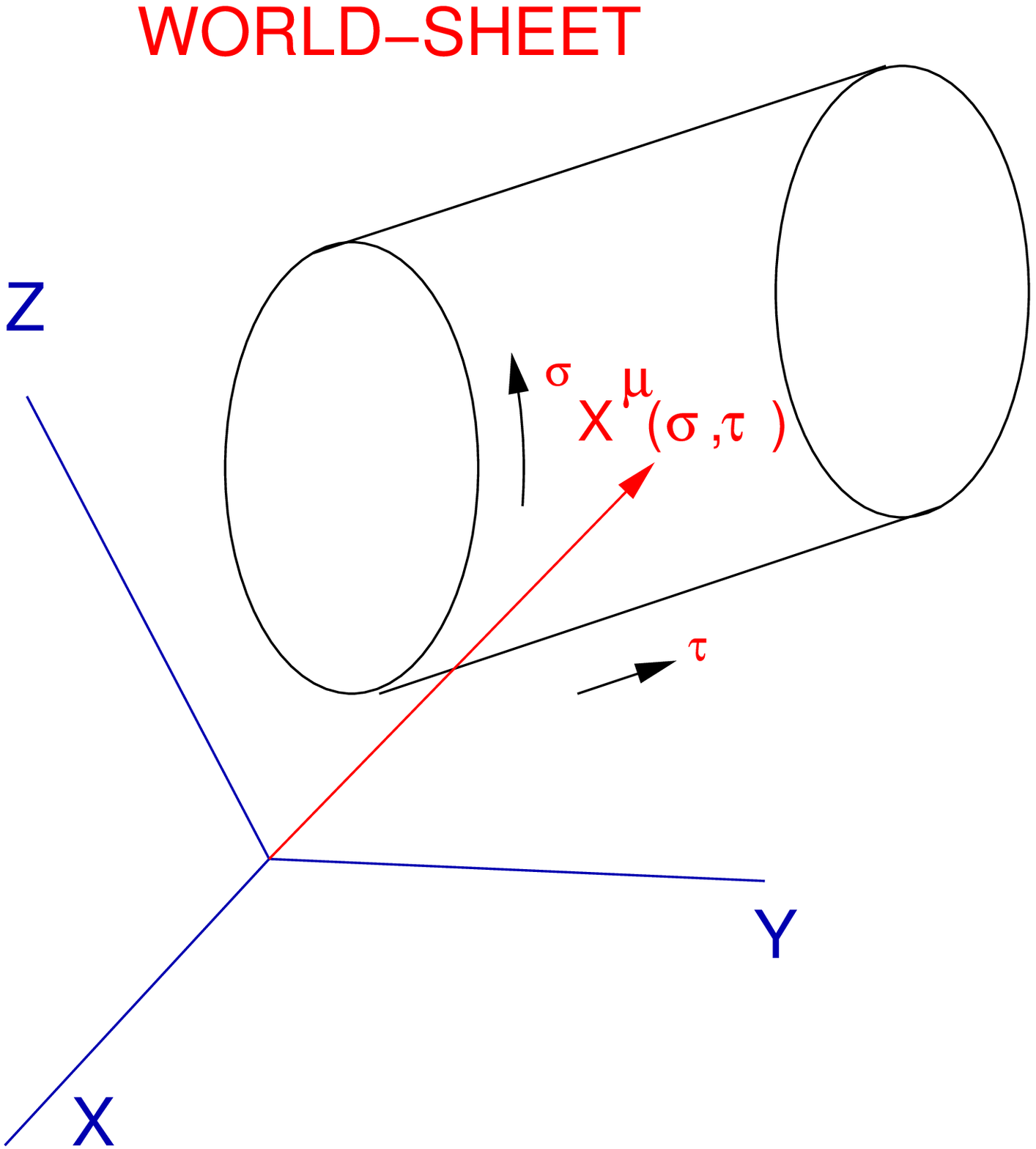,height=2.5in}
\caption{Left panel: The world sheet of an OPEN string
propagating in the space-time.  Right panel: The world sheet of a
CLOSED string.
\label{fig2}}
\end{figure}
\end{center}

For a consistent quantum theory description of gravity we must abandon
the concept of particle and instead introduce that of
string~\cite{Witten}. Strings and in particular superstrings will
lead, as we will see, to the existence of extra dimensions. In fact a
string is a generalization of a ``point particle''. While the latter
is described, when propagating in the space-time, by a world line
$X^\mu=X^\mu(\tau)$, where $\tau$ is the proper time of the particle,
a string is a one-dimensional spatially extended object that, when
propagating in D space-time dimensions, spans a world sheet $
X^\mu=X^\mu(\xi^\alpha), \quad \xi^\alpha=(\tau,\sigma)$, where now
there are a proper time $\tau$ and a proper space $\sigma$. Strings
can be open or closed in the proper space direction $\sigma$, as shown
in Fig.~\ref{fig2}.

The cancellation of the conformal anomaly leads to a critical value of
the space-time dimensionality. In the case of supersymmetric strings
it is given by $D=10$ ($D=11$ for
M-theory~\cite{Horava:1996ma}). There are then six extra dimensions on
top of the four flat ones that we experience. The six extra dimensions
must be compactified and the size of their radii is constrained by
experimental data, as we will see. The existence of extra
(compactified) spatial dimensions is the first remarkable feature of
string theories.

\section{Strings and branes}

The second great feature of string theories is that they predict the
existence of subsurfaces of the whole space or ``branes''. In
particular $Dp$-branes are $(p+1)$-dimensional subsurfaces where open
strings can end. Extra dimensions can then be either LONGITUDINAL or
TRANSVERSE to the branes. Technically speaking longitudinal directions
are those where the string has Neumann boundary conditions while on
transverse directions the string has Dirichlett boundary
conditions. In other words open string ends can move along
longitudinal (Neumann) coordinates and therefore they have
Kaluza-Klein (KK) modes with respect to them while they are fixed at
transverse (Dirichlett) coordinates and have winding modes ($\omega$)
depending on the number of times the string winds around the
transverse (compact) coordinate.  Closed strings propagate in the
bulk.  These situations are illustrated in Fig.~\ref{fig3}

\begin{center}
\hskip 2cm
\begin{figure}[htb]
\psfig{figure=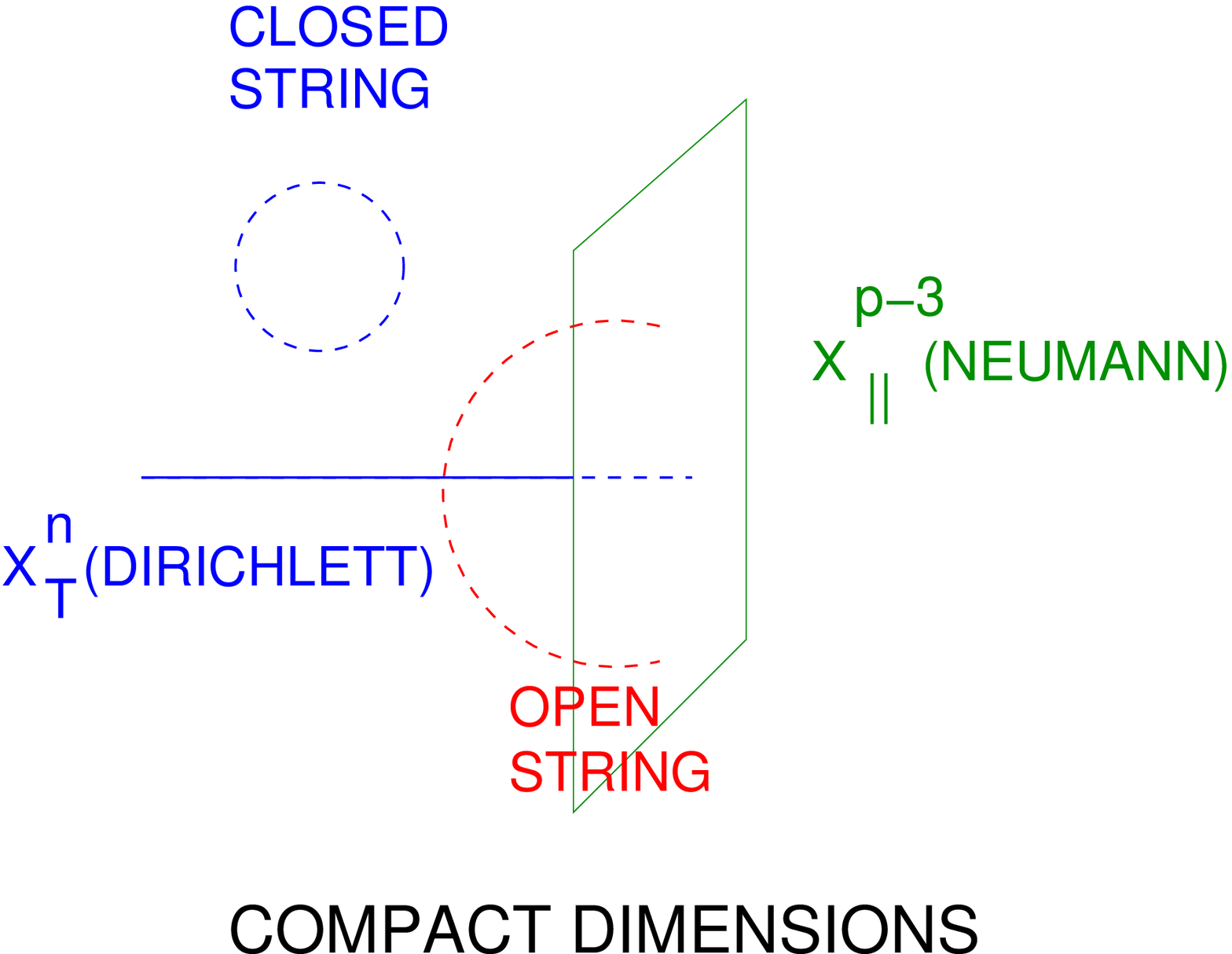,height=2.5in}\hfill\vspace{.5cm}
\psfig{figure=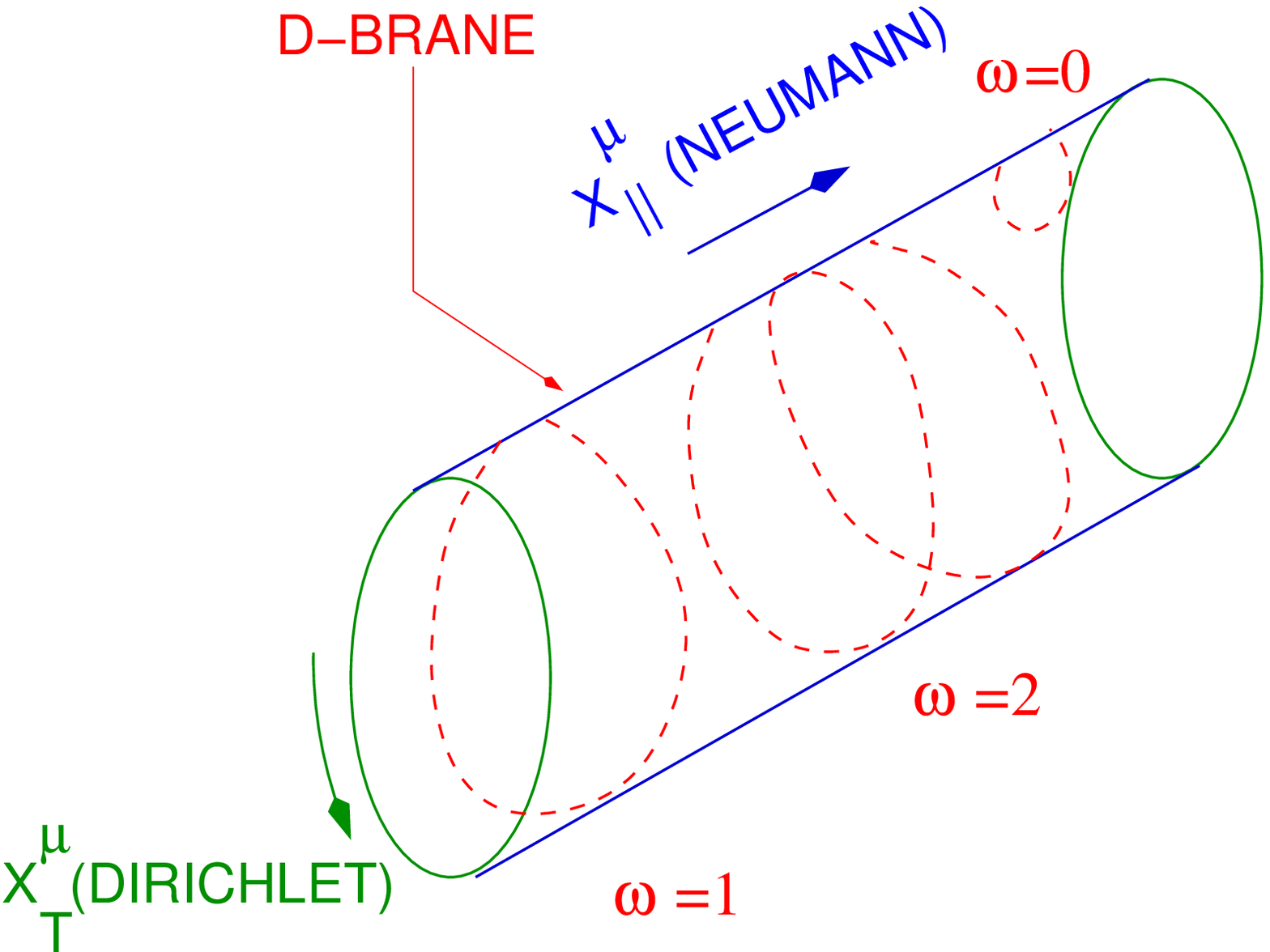,height=2in}
\caption{Open and closed strings propagating in extra
dimensions. Closed strings propagate in the bulk (left panel). Open
strings propagate with ends at $X_T=$constant for different windings
(right panel) while they depend on $X_L\equiv X_{||}$.
\label{fig3}}
\end{figure}
\end{center}

To make contact with experiments we will describe the different string
theories from the point of view of effective field theories. The
parameters are

$\bullet$ $String\ scale\quad {\Rightarrow}\quad M_s=\ell_s^{-1}$,
where $\alpha^\prime\sim 1/M_s^2$ is the string tension.

$\bullet$ $ String\ coupling\quad { \Rightarrow}\quad
\lambda_s=e^{\langle\phi\rangle}$, where $\phi$ is the dilaton field.

$\bullet$ COMPACTIFICATION $10D\Rightarrow 4D\otimes V_6$

$\bullet$ $Planck\ scale\quad { \Rightarrow}\quad  M_P=\ell_P^{-1}$

$\bullet$ $Gauge\ coupling\quad { \Rightarrow}\quad g$

\noindent
The massless modes of open and closed strings describe the physical
degrees of freedom of our four-dimensional effective theories. In
particular

$\bullet$ Closed strings describe gravity

$\bullet$ Open strings with ends bounded to propagate on
Dp-branes describe gauge interactions.

\noindent
The relationship between the string parameters, string scale ($M_s$)
and the string coupling ($\lambda_s$), and the four-dimensional
parameters, Planck scale ($M_{P\ell}$) and unified gauge coupling
($g$) is provided by the compactification from ten to four dimensions.
The six internal compact dimensions are [p-3] (longitudinal) and
[n=(9-p)](transverse). The ten-dimensional effective action can be
written as (the different powers of string parameters are easily
obtained)
\begin{equation}
S=\int_{{bulk}}d^{10}x\frac{1}{\lambda_s^2}
\ell_s^{-8}\mathcal{{R}}+
\int_{{brane}}d^{p+1}x\frac{1}{\lambda_s}\ell_s^{3-p}{F^2}
\end{equation}

Upon compactification of extra dimensions the following relations
between string and four-dimensional parameters follow
\begin{equation}
\frac{1}{\ell_P^2}=\frac{V_{L} V_T}{\lambda_s^2\ell_s^8}\qquad
\frac{1}{g^2}=\frac{V_{L}}{\lambda_s\ell_s^{p-3}}
\end{equation}
and rescaling the longitudinal volume as $ V_{L}\equiv
v_{L}\ell_s^{p-3} $ one gets the final relation
\begin{equation}
M_P^2=\frac{1}{g^4 v_{L}}M_s^{2+n} R_T^n \qquad
 \lambda_s=g^2 v_{L}
\end{equation}
Defining now the (4+n) Planck scale $M_*$ as $M_{*}^{2+n}=
M_s^{2+n}/g^4 v_{L}$ one can finally write the ADD relation as
\begin{equation}
M_P^2=M_*^{2+n} R_T^n
\label{ADD}
\end{equation}

\section{Experimental detection}

If $M_s\sim M_*\sim$ few TeV string excitations can be at reach at LHC.
On the other hand the ADD relation can ``explain'' the weakness of
gravitational interactions by the size of extra-large transverse
dimensions $1/R_T\ll 1$ TeV. Both transverse and longitudinal
dimensions (i.~e.~their Kaluza-Klein excitations) can be at reach at
experiments:

$\bullet$ Kaluza-Klein excitations of transverse dimensions $R_T$
can affect gravitational (and collider) experiments.

$\bullet$ Kaluza-Klein excitations of longitudinal dimensions with
size $1/R_{L}\sim$ few TeV can be detected at LHC.

\subsection{Transverse dimensions}
Only gravity propagates in transverse extra dimensions. Therefore
these dimensions can be extra-large.  Let $R_T$ be the (common) radius
of transverse dimensions where gravity propagates.  The ADD relation
(\ref{ADD}) relates the scales of quantum gravity in the higher
dimensional theory and the transverse radius. For instance for a value
of the fundamental scale $M_*=2$ TeV the value of the common
transverse radius $R_T$ depends on the different possible
dimensionalities ($n$) of the transverse space. This is illustrated in
Table~\ref{table1}.

\begin{table}[htb]
\begin{center}
\caption{The prediction of the transverse radius for different
dimensions of the transverse space}
\vspace{.4cm}
\label{table1}
\begin {tabular}{|c|c|c|c|}\hline
n & 1 & 2 & 6\\\hline
$R_T$ & $10^7$ Km& { 0.2 mm} & {0.1 fm}\\ \hline
&{\sc excluded}& {\sc  barely}& {\sc  consistent }\\
&&{\sc  consistent} &\\
\hline
\end{tabular}
\end{center}
\end{table}

\noindent
As we can see from Table~\ref{table1} the case of $n=1$ transverse
dimension is excluded while the case of $n=2$ is barely consistent
with table-top gravitational experiments.

\subsubsection{-- Gravitational experiments}
Extra dimensions produce deviations from Newton's law. They can be
parametrized in the following way:
\begin{equation}
V(r)=\left\{
\begin{array}{l}
-\frac{G_N}{r}\left(1+\alpha e^{-r/R_T}+\cdots \right),\
 r\geq R_T\\
-\frac{G_N^{(4+n)}}{r^{1+n}},\ r\ll R_T
\end{array}\right.
\end{equation}
Present gravitational experiments place the lower bounds on $R_T$ in the
sub-millimeter region. Some recent experimental results can be found
in Ref.~\cite{tabletop}. These bounds appear in the typical way shown
in Fig.~\ref{fig4}.
\begin{center}
\hskip 2cm
\begin{figure}[htb]
\hspace{1.2cm}\psfig{figure=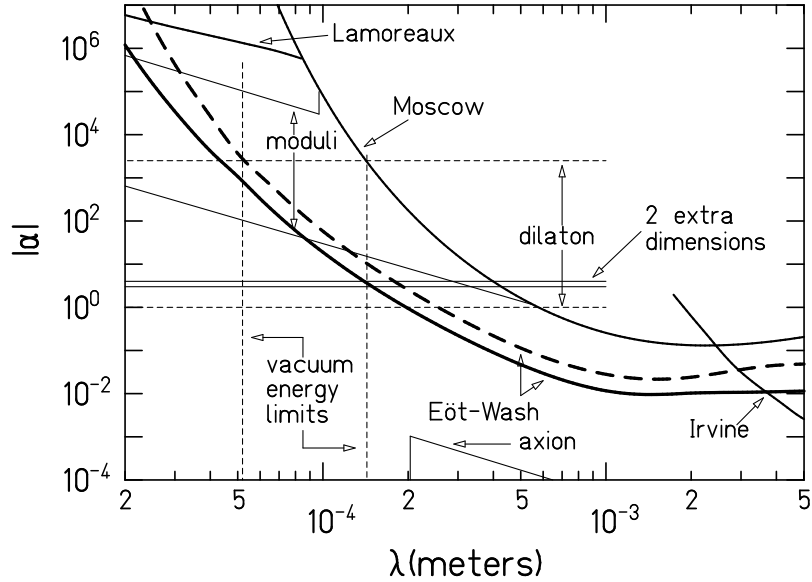,height=3.5in}
\caption{Limit on non-Newtonian forces at short distances as a
function of their range $\lambda$ and their strength relative to
gravity $\alpha$. Present bounds are in the sub-millimeter region.
\label{fig4}}
\end{figure}
\end{center}

\subsubsection{-- Collider signatures}
Transverse extra dimensions can also be detected in high-energy
colliders. The signatures are based on missing energy in reactions
corresponding to the production of {KK-gravitons} in the {bulk}. For
instance
\begin{equation}
e^+e^-\longrightarrow \gamma {\displaystyle \sum_k}G^{(k)}
\end{equation}
Every single graviton couples gravitationally {$\sim 1/M_P^2$} but the
large amount of gravitons {cancels} (using ADD relation) the
$M_{P\ell}^2$ dependence
\begin{equation}
\sigma\sim s^{n/2}/M_{*}^{n+2}
\end{equation}
The 95\% confidence limits on $R_T$ [cm] and $M_{*}$ [GeV] have been
computed in Ref.~\cite{Mirabelli:1998rt}, where $R_T$ and $M_{*}$ are
related by the ADD relation. For present or past accelerators the
results are (depending on the dimensionality of the extra space) given
in Table~\ref{table2}.

\begin{table}[htb]
\caption{Bounds on $M_*$ [GeV] and $R_T$ [cm] from LEP2 and Tevatron }
\vspace{.4cm}
\label{table2}
\begin{center}
\begin{tabular}{|l  | r |  r| } \hline
  {\sc Collider} &  $R_T$ / $M_{*}$ ($n=2$) & $R_T$ / $M_{*}$ ($n=6$) \\
           \hline\hline
  
   {\sc LEP2} & $4.8\times 10^{-2}$ / 1200 & $6.9 \times
                                       10^{-12}$ / 520 \\ \hline 
                                       {\sc Tevatron} & $11.0 \times
                                       10^{-2}$ / 750 & $5.8 \times
                                       10^{-12}$ / 610 \\
\hline
\end{tabular}
\end{center}
\end{table}

\noindent For future accelerators (as the LHC and the ILC) the bounds
obtained in~\cite{Mirabelli:1998rt} range for $M_*$, between 8 TeV
(for $n=2$) and 3 TeV (for $n=6$), and for $R_T$ between $10^{-3}$ cm
(for $n=2$) and $10^{-13}$ cm (for $n=6$).

\subsection{Longitudinal directions}

Longitudinal directions are those where the Standard Model (SM) fields
can propagate. In this picture the SM fields propagate in a {brane},
with p-3 longitudinal dimensions wrapped on a {compact} space
(orbifold) with 4D boundaries at the fixed points. The cartoon of the
brane looks like in Fig.~\ref{fig5}.

\begin{center}
\hskip 2cm
\begin{figure}[htb]\hspace{3.5cm}
\psfig{figure=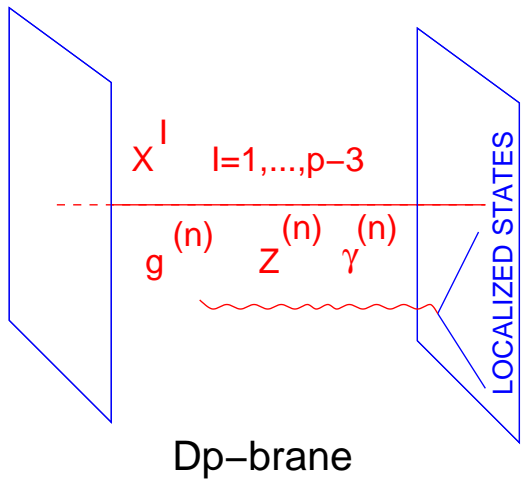,height=2.5in}\vspace{.5cm}
\caption{A $p$-brane where SM fields propagate. We will consider that
the gauge sector is propagating in the bulk of the extra dimensions
while the quark and lepton sector is localized at the orbifold fixed
points.
\label{fig5}}
\end{figure}
\end{center}

\subsubsection{-- Indirect detection}

\begin{center}
\begin{figure}[htb]\hspace{1.5cm}
\psfig{figure=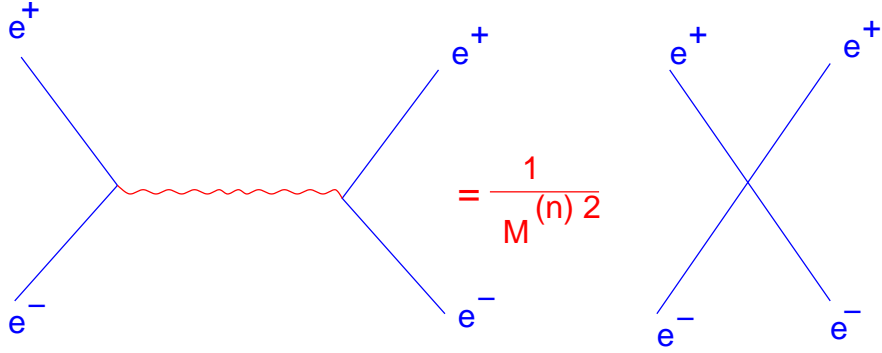,height=1.8in}\vspace{.5cm}
\caption{Fermions localized on the brane couple to KK modes and give rise to
an effective six-dimensional operator.
\label{fig6}}
\end{figure}
\end{center}

The indirect detection of longitudinal extra dimensions is based on
the modification of EW observables $(M_W,\, \Gamma_{\ell\ell},\,
\Gamma_{had},\, A_{FB}^\ell,\, Q_W,\,\dots)$ by the exchange of
KK-modes $Z^{(n)}$, $\gamma^{(n)}$, as it is shown in the picture of
Fig.~\ref{fig6}.

\begin{center}
\begin{figure}[htb]
\psfig{figure=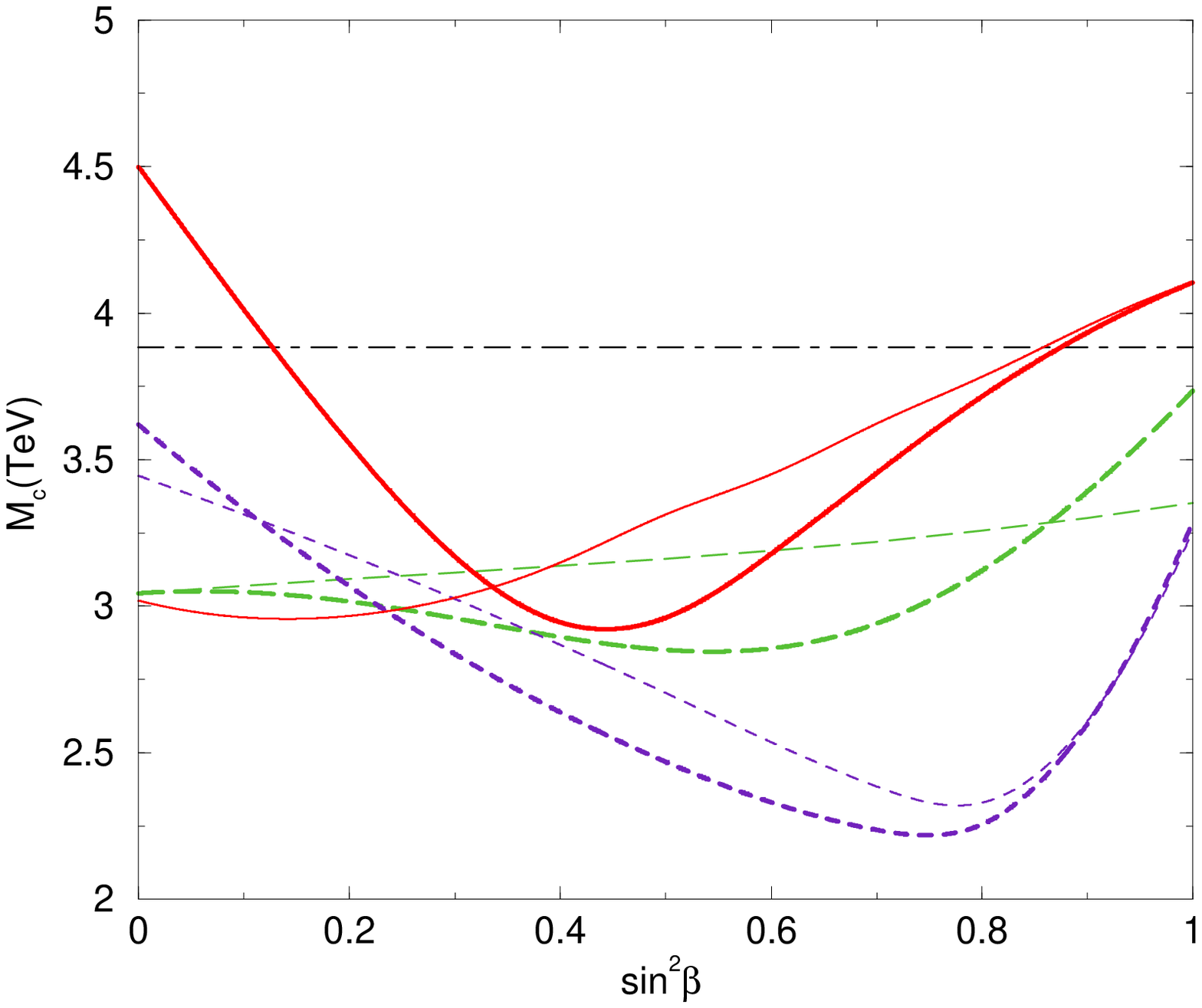,height=2.5in}
\psfig{figure=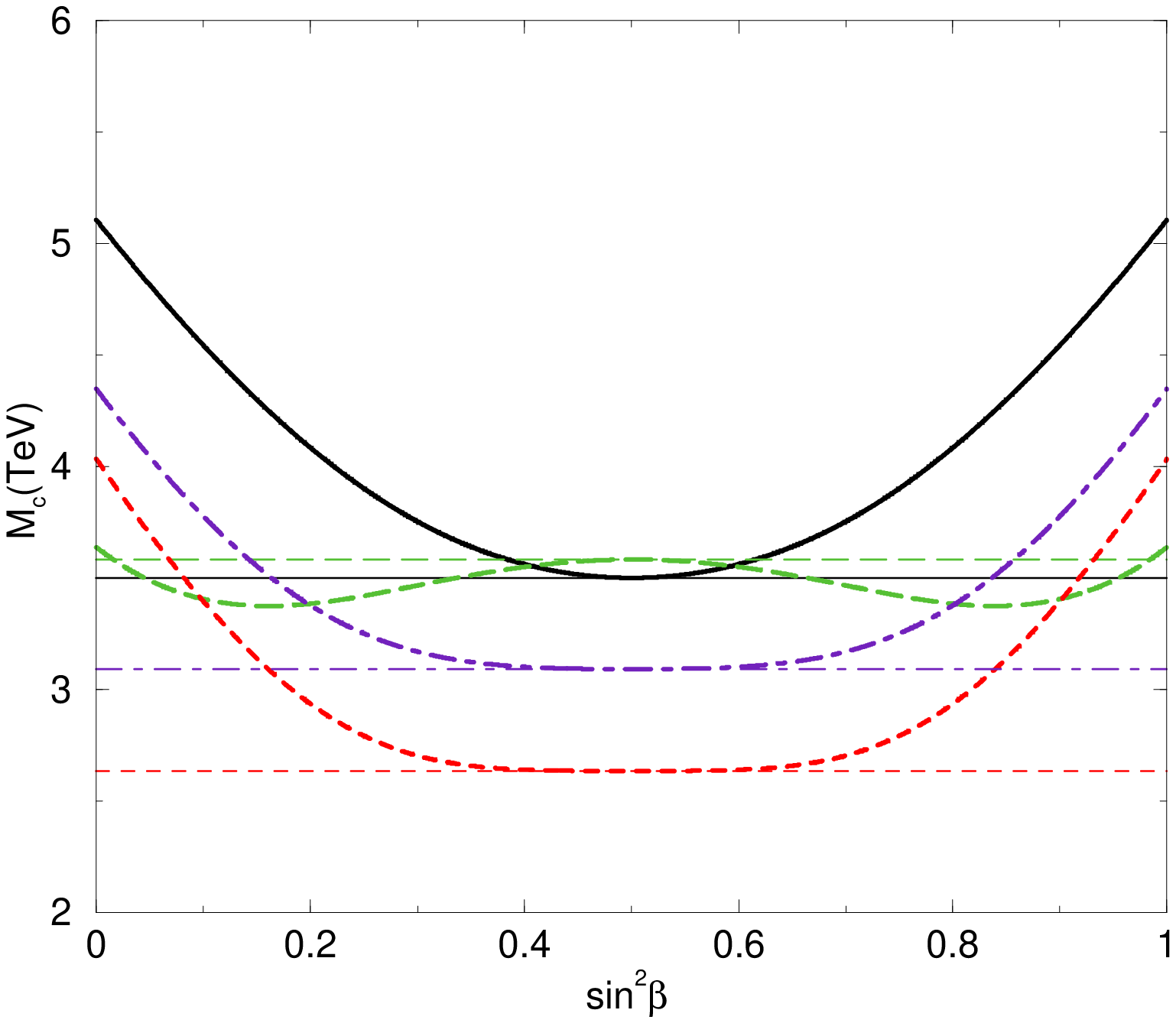,height=2.5in}
\caption{Lower bounds on $M_c$ in five-dimensional models from
electroweak precision tests.
\label{fig7}}
\end{figure}
\end{center}

Electroweak precision tests provide bounds on the compactification
scale $M_c=1/R$ that are model dependent and in all cases of order a
few TeV. To give an idea of the model dependence of the result we
present in Fig.~\ref{fig7} the bounds on $M_c$ as a function of
$\sin^2\beta$ (where $\tan\beta=v_2/v_1$) for the different models
analyzed in Ref.~\cite{DPQ}. For more details see Ref.~\cite{DPQ}.

\subsubsection{-- Direct detection}

\begin{center}
\begin{figure}[htb]\hspace{3.cm}
\psfig{figure=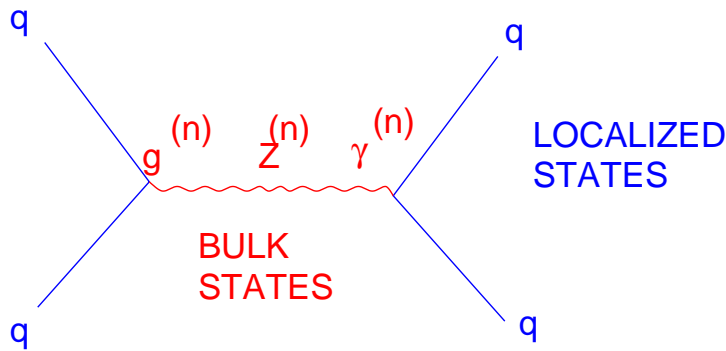,height=1.8in}
\caption{Drell-Yan process through exchange of the KK mode of a gluon,
electroweak gauge boson or photon.
\label{fig8}}
\end{figure}
\end{center}

Direct detection is based on production of KK modes from $f\bar f$ and
subsequent decay into a pair of fermions. For instance in hadron
colliders (as Tevatron or LHC) it will proceed through Drell-Yan
processes as that presented in Fig.~\ref{fig8}.

\begin{center}
\begin{figure}[h]
\hspace{1cm}\psfig{figure=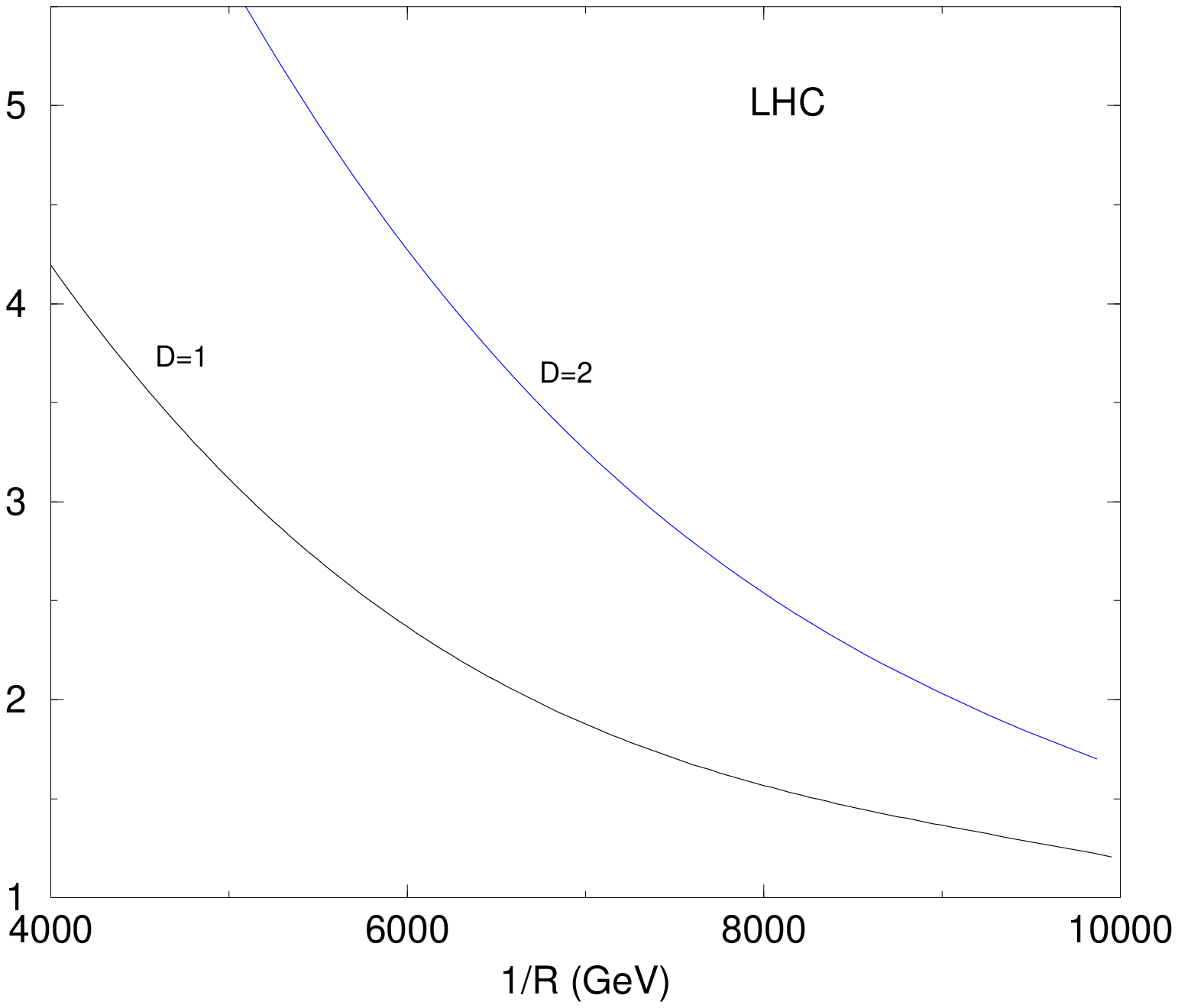,height=2.2in}
\psfig{figure=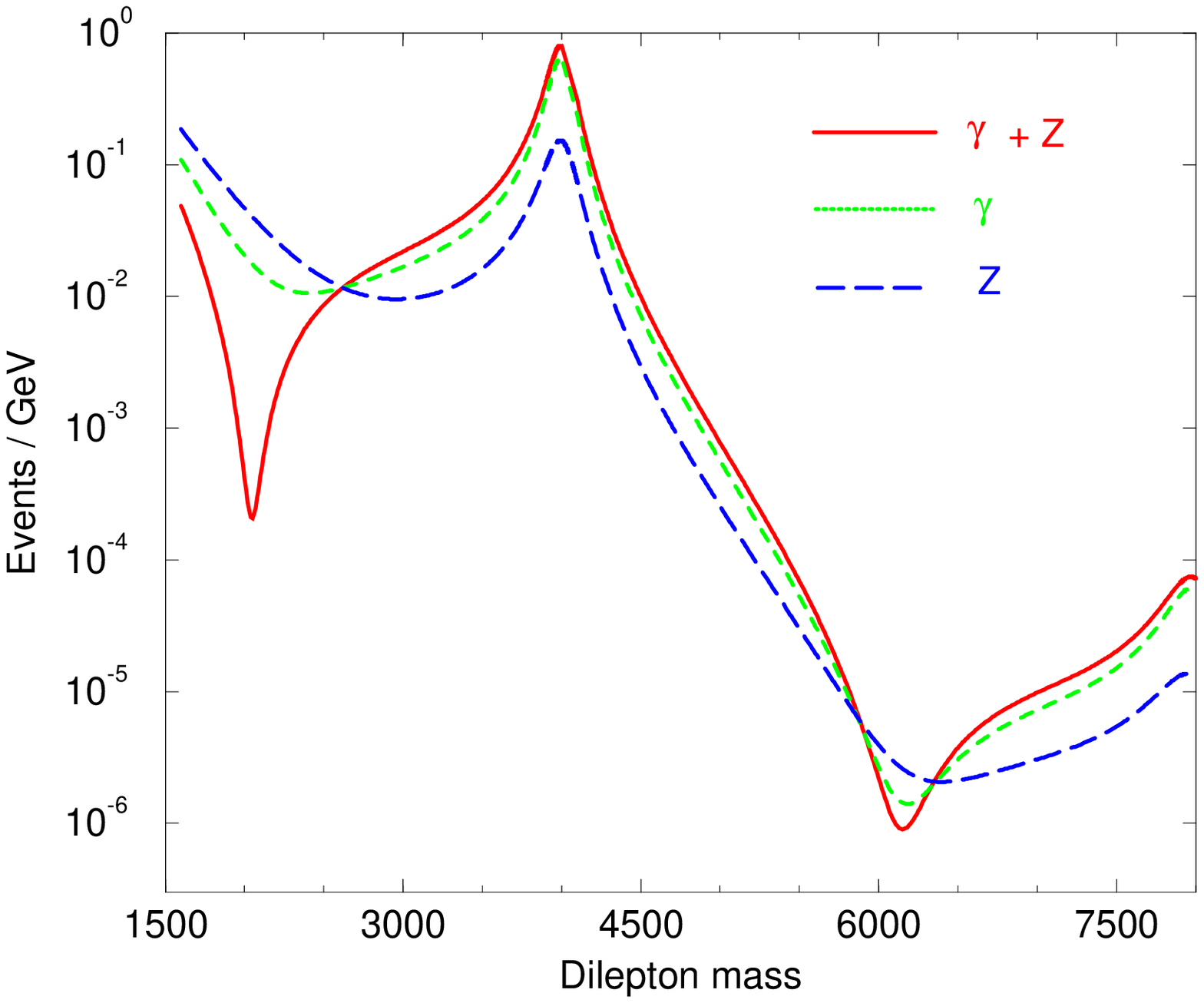,height=2.5in}
\caption{Left panel: Excess of lepton pair events at LHC for dilepton
invariant mass above 400 GeV from $D=1,2$ extra dimensions. Right
panel: Production at LHC of $\gamma$ and $Z$ Kaluza-Klein excitations.
\label{fig9}}
\end{figure}
\end{center}

The direct production of KK-modes through Drell-Yan processes will in
turn produce two effects. One is that it will increase the number of
events with respect to the Standard Model prediction. For instance for
the LHC collider and a given number of extra dimensions ($D=1,2$) one
can predict the excess in the number of events $(N_N^{\rm
SM})/\sqrt{N^{\rm SM}}$ as a function of the longitudinal radius $R_L$
as in the left panel of Fig.~\ref{fig9}~\cite{Antoniadis:1999bq}. One
can see that, depending on the number of extra dimensions, one can
reach 9~TeV for the 95 \% confidence level bounds. The other effect is
when the resonance is detected and produces a bump on the invariant
mass of the final states. This one is the ``golden channel'' to
discover an extra gauge boson. In the case of KK excitations, and
depending on the compactification radius, one can even detect the
first two-KK modes for the case of one extra dimension as shown in the
right panel of Fig.~\ref{fig9}.

We will conclude this section by summarizing in Table~\ref{table3} the
largest compactification scales that can be probed at 95 \% confidence
level for one extra dimension and different colliders.
\begin{table}[htb]
\caption{Highest compactification scales at reach for one extra
dimension.}
\vspace{.4cm}
\label{table3}
\begin{center}
\begin{tabular}{|l | c | c|c|c|c| } \hline
  {\sc Collider} &  LEP2 & ILC-500 & ILC-1000& Tevatron& LHC \\
           \hline
  
  $M_c$ (TeV) & 1.9&8&13&1.2&6.7\\ 
\hline
\end{tabular}
\end{center}
\end{table}


\section{New theoretical ideas from extra dimensions}
Extra dimensions also exhibit the feature of providing {NEW} solutions
to {OLD} problems in Particle Physics. These solutions mainly concern
the problem of supersymmetry breaking as well as electroweak symmetry
breaking. In this section we will cover a sample of possibilities and
will leave aside others (as e.~g. Higgsless models). Our selection is
biased by our own experience and does not reflect at all any
scientific preference.

\subsection{Supersymmetry breaking}

$\bullet$ Supersymmetry can be broken on the fixed points by orbifold
boundary conditions~\cite{Dixon:1985jw}: the number of supersymmetries
on the branes is halved. For instance in the case of {one} or {two}
extra dimensions compactified on orbifolds the number of
supersymmetries of zero modes is

\begin{center}
\fbox{\parbox{3.5cm}{
{ $N=2\quad {\Rightarrow}\quad N=1$}
}}
\end{center}

\noindent $\bullet$ Supersymmetry can be further broken by {twisted
boundary conditions}: the so-called {Scherk-Schwarz
breaking~\cite{SS}}. Then in the effective four-dimensional theory

\begin{center}
\fbox{\parbox{3.5cm}{
{$N=1\quad {\Rightarrow}\quad N=0$}
}}
\end{center}

\noindent $\bullet$ In particular if {gauginos} (and squarks)
propagate in the bulk of extra dimensions they acquire masses
proportional to the { compactification scale}

\begin{center}
\fbox{\parbox{2.5cm}{$M_{1/2}\sim 1/R$
}}
\end {center}

\subsection{Gauge-Higgs unification}

It is possible to unify the Higgs and gauge sectors by using for the
former the extra dimensional components of the higher-dimensional
gauge bosons: this is called gauge-Higgs  unification

\begin{center}
\fbox{\parbox{5cm}{ {$A_\mu^\alpha={\rm gauge\ bosons},\\ 
A_i^\alpha={\rm Higgses};\ \ { \alpha\in {\rm Adjoint}}$} }}
\end {center}

The two main features of the gauge-Higgs unification are:

\noindent $\bullet$ It requires to enlarge the SM gauge group
e.~g.~\cite{NJP}

\begin{center}
\fbox{\parbox{5cm}{
{$SU(3)\to SU(2)\otimes U(1)$}
}}
\end {center}

\noindent $\bullet$ This is an alternative solution to the {hierarchy
problem} because the {extra dimensional gauge theory} protects the
Higgs mass from quadratic divergences~\cite{NJP}.

\noindent $\bullet$ Electroweak breaking proceeds via the Hosotani
mechanism~\cite{Hosotani} (Wilson line). For instance in the simplest
model~\cite{NJP}

\begin{center}
 {$ A_\mu^{SU(2)\otimes U(1)}$= SM gauge bosons} 
\end{center}

\begin{center}
{$  A_i^{SU(3)/SU(2)\otimes U(1)}$= SM Higgs bosons}
\end{center}

\noindent $\bullet$ The Higgs mass parameter becomes tachyonic radiatively

\begin{center}
\fbox{\parbox{5cm}{
{ $m_H^2\sim -({\rm loop\ factor})\ 1/R^2$}
}}
\end {center}

\noindent $\bullet$ It is difficult to obtain realistic models:
quadratically divergent localized tadpoles can appear, Higgs mass
requires more than five dimensions, fermion masses are difficult to
obtain,~\dots

\subsection{Higgs Electroweak Symmetry Breaking}

By using the Scherk-Schwarz mechanism to break supersymmetry and the
matter localized on the four-dimensional boundary (gaugino mediated
supersymmetry breaking)~\cite{Pomarol:1998sd} one can obtain a
realistic and very peculiar model of electroweak symmetry
breaking~\cite{David}. This model has the following nice features:

\noindent $\bullet$
Models are of "no-scale" type and then no anomaly mediated
supersymmetry breaking occurs.

\noindent $\bullet$
No one-loop quadratic or linear sensitivity on the cutoff $\Lambda$ of Higgs
masses. 

\noindent $\bullet$
Gauginos are the heaviest supersymmetric particles (they are
in the TeV or multi-TeV region).

\noindent $\bullet$
EWSB is triggered by tachyonic tree-level masses and two-loop
radiative corrections.

\noindent $\bullet$ Squarks and sleptons acquire radiative masses from
gluinos and electroweak gauginos, respectively. In fact there is a
``fixed'' mass relation that can be considered as the smoking gun of
these models. It is given by
\begin{equation}
(m_{\tilde
q_L},\,m_{\tilde u_R},\,m_{\tilde d_R},\,m_{\tilde \ell_L},\,m_{\tilde
e_R}) \simeq \\ (0.110,\,0.103,\,0.102,\,0.042,\,0.025)\sin\pi\omega /R
\end{equation}
where the SS parameter is $\omega=M_{\tilde g} R$.

\noindent $\bullet$
Charged and neutral Higgsinos are almost
degenerate with mass splittings $\sim 1$ GeV.

\noindent $\bullet$
Fine-tuning of MSSM alleviated.

\noindent $\bullet$
The LSP is a neutralino which is a good candidate to Dark Matter.

\section{Conclusions}

We will conclude with a few obvious observations:

\begin{itemize}
\item
Large extra dimensions are well motivated theoretically.
\item
 Large extra dimensions and low scale quantum gravity effects are at reach
at present (Tevatron) and future (LHC) colliders.
\item
Large extra dimensions have unambiguous experimental signatures.
\item
Large extra dimensions can also help to solve theoretical Particle
Physics problems (electroweak breaking, hierarchy, flavor,...).
\item
If extra dimensions are found at LHC it would possibly constitute the
most important revolution in the History of Particle Physics!

\end{itemize}
\section*{Acknowledgments}
\noindent This work was partly supported by CICYT, Spain, under
contracts FPA 2004-02012 and FPA 2005-02211.

\end{document}